\documentclass[aps,prl,twocolumn,showpacs,superscriptaddress]{revtex4}

\usepackage[dvips]{graphicx} 
\usepackage{natbib}

\newcommand{\srI}{Sr$_{2}$RuO$_{4}$}

\newcommand{\srII}{Sr$_{3}$Ru$_{2}$O$_{7}$}
\newcommand{\caII}{Ca$_{3}$Ru$_{2}$O$_{7}$}

\newcommand{\ie}{\textit{i.e.}}

\begin{document}

\title{Fermi surface and van Hove singularities in the itinerant metamagnet Sr$_3$Ru$_2$O$_7$}


\author{A. Tamai}
\email[]{anna.tamai@st-andrews.ac.uk}
\affiliation{Scottish Universities Physics Alliance, School of Physics and Astronomy, University of St Andrews, North Haugh, St. Andrews, Fife KY16 9SS, UK}
\author{M.P. Allan}
\affiliation{Scottish Universities Physics Alliance, School of Physics and Astronomy, University of St Andrews, North Haugh, St. Andrews, Fife KY16 9SS, UK}
\author{J.F. Mercure}
\affiliation{Scottish Universities Physics Alliance, School of Physics and Astronomy, University of St Andrews, North Haugh, St. Andrews, Fife KY16 9SS, UK}
\author{W. Meevasana}
\affiliation{Departments of Applied Physics, Physics, and Stanford Synchrotron Radiation Laboratory, Stanford University, Stanford, California 94305, USA}
\author{R. Dunkel}
\affiliation{Departments of Applied Physics, Physics, and Stanford Synchrotron Radiation Laboratory, Stanford University, Stanford, California 94305, USA}
\author{D.H. Lu}
\affiliation{Departments of Applied Physics, Physics, and Stanford Synchrotron Radiation Laboratory, Stanford University, Stanford, California 94305, USA}
\author{R.S. Perry}
\affiliation{Scottish Universities Physics Alliance, School of Physics and Centre for Science at Extreme Conditions, The University of Edinburgh, Mayfield Road, Edinburgh, EH9 3JZ, UK}
\author{A.P. Mackenzie}
\affiliation{Scottish Universities Physics Alliance, School of Physics and Astronomy, University of St Andrews, North Haugh, St. Andrews, Fife KY16 9SS, UK}

\author{D.J. Singh}
\affiliation{Materials Science and Technology Division, Oak Ridge National Laboratory, Oak Ridge, TN 37831-6114, USA}
\author{Z.-X. Shen}
\affiliation{Departments of Applied Physics, Physics, and Stanford Synchrotron Radiation Laboratory, Stanford University, Stanford, California 94305, USA}
\author{F. Baumberger}
\affiliation{Scottish Universities Physics Alliance, School of Physics and Astronomy, University of St Andrews, North Haugh, St. Andrews, Fife KY16 9SS, UK}


\date{\today}

\begin{abstract}
The low-energy electronic structure of the itinerant metamagnet \srII{} is investigated by angle resolved photoemission and density functional calculations. We find well--defined quasiparticle bands with resolution limited line widths and Fermi velocities up to an order of magnitude lower than in single layer \srI. The complete topography, the cyclotron masses and the orbital character of the Fermi surface are determined, in agreement with bulk sensitive de Haas -- van Alphen measurements. 
An analysis of the $d_{xy}$ band dispersion reveals a complex density of states with van Hove singularities (vHs) near the Fermi level; a situation which is favorable for magnetic instabilities.
\end{abstract}


\maketitle

Quantum criticality in correlated electron systems continues to attract widespread attention. One reason is that systems near a quantum critical point are highly susceptible to novel ordered phases, the observation and characterization of which promises new insight in the behavior of strongly interacting systems \cite{sac00, bor07ea}. Possibly even more exciting is the prospect that fluctuations associated with a quantum critical point could dominate the phase diagram of topical materials, including the cuprate superconductors up to high temperatures \cite{var97}. Evidence for quantum critical points, reached by tuning pressure, chemical composition or magnetic field, has been observed in a variety of materials comprising simple metals \cite{yeh02}, heavy fermion intermetallics \cite{ste01} or transition metal oxides such as \srII, the subject of this study \cite{gri01, bor07ea}.

Criticality in \srII{} is associated with a metamagnetic transition (super--linear rise in magnetization) of the itinerant electron system in applied field. In the ground state \srII{} is a paramagnetic Fermi liquid with strongly enhanced quasiparticle masses. Its electronic specific heat value of $\gamma = 110$~mJ/molRuK$^2$ is among the highest in any oxide and the large magnetic susceptibility indicates a substantial Stoner enhancement \cite{ike00}. The Fermi liquid region of the phase diagram extends up to 10 -- 15 ~K in zero field and is continuously suppressed towards zero temperature upon approaching the critical field of $B \approx 8$~T (for $B\parallel c$) \cite{gri01}. 
In the vicinity of the putative quantum critical end point non--Fermi--liquid behavior has been observed in various macroscopic quantities including specific heat, resistivity and thermal expansion and has been described on the basis of phenomenological models \cite{gri01, per01ea, geg06, mil02}.

On the other hand, little is known about the microscopic origin of the metamagnetism. Theoretical work suggests that the phase diagram of \srII{} may be understood from peculiarities in the band structure 
causing either a local minimum in the density of states (DOS) at the Fermi level $\rho(E_F)$ or a sharp increase in DOS over the minute energy scale of the Zeeman splitting ($\approx 1$~meV for $B=10$~T) \cite{woh62,bin04,hon05}.
Density functional calculations for \srII{} show a number of sharp features in the DOS \cite{sin01} and have been invoked to support a band structure based model of metamagnetism, although their precision hardly reaches the 1~meV scale. 
Thus, experimental information on the evolution of the DOS near the Fermi level is key to advancing the field.
A recent spectroscopic STM study detected peaks in the tunneling conductivity around $\pm 4$~meV, which might be interpreted as a set of van Hove singularities in the DOS \cite{iwa07ea}. Published photoemission studies on the other hand did not report unusual features on the relevant energy scale \cite{puc98b, aiu04ea}.

In this paper, we present a comprehensive high-resolution k--space mapping of the quasiparticle band structure in \srII{} by means of angle-resolved photoemission. 
We characterize the $d_{xy}$ DOS in the vicinity of the chemical potential: we identify two peaks in $\rho(\epsilon)$ at $-4$~meV and $-1$~meV and discuss their relevance for metamagnetism.
\begin{figure*}[tb]
\includegraphics[width=0.97\textwidth]{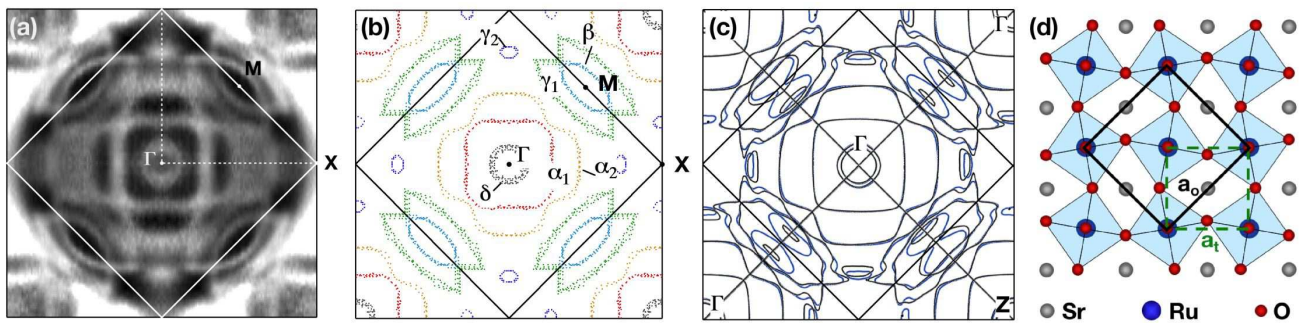}
\caption{\label{f1} Fermi surface of \srII. (a) shows the experimental data taken in the first quadrant of the larger tetragonal Brillouin zone and symmetrized with respect to the Ru--Ru nearest neighbor direction. $X$ denotes the surface projection of R and $M$ that of a midpoint between two $\Gamma$ points ($Z$ or $S$). (b) Fermi surface contours extracted from the data shown in panel (a). (c) LDA calculation for the basal plane ($k_z = 0$, black) and  midplane ($k_z = 1/4$, blue). (d) Schematic structure of a single RuO$_2$ plane illustrating the unit-cell doubling due to a $6.8^{\circ}$ rotation of the RuO$_6$ octahedra \cite{sha00}. $a_t$ denotes the Ru--Ru nearest neighbor distance and $a_o$ the in--plane lattice constant of the orthorhombic unit cell, respectively.}
\end{figure*}
%
%

Photoemission experiments were performed with a monochromatized He--discharge lamp and a Scienta SES2002 analyzer using an energy and angular resolution of 4.5~meV / 0.3$^{\circ}$ full width at half maximum (FWHM). Additional data at various excitation energies were taken at beamline V-4 of SSRL using a Scienta R4000 analyzer with a combined resolution set to 8.0~meV / 0.3$^{\circ}$. All data shown in this paper were measured with 21.2~eV photons at T = 9~K and a pressure $< 5\cdot 10^{-11}$~mbar. High purity single crystals of \srII{} were grown in an image furnace \cite{per04a}. The detection of sizable quantum oscillations in all samples used for this study indicates residual in--plane resistivities $<0.5$~$\mu\Omega$cm. 
Band structure calculations in the local density approximation (LDA) were performed using the general potential linearized augmented planewave method \cite{singh} with well converged basis sets and zone samplings.  Including spin--orbit interaction  was found to improve the agreement with experiment. The Fermi surface is based on first principles calculations at 850 $k$--points in the irreducible $1/8$~wedge of the zone using the experimental crystal structure \cite{sha00}.

\begin{figure}[!b]
\includegraphics[width=0.46\textwidth]{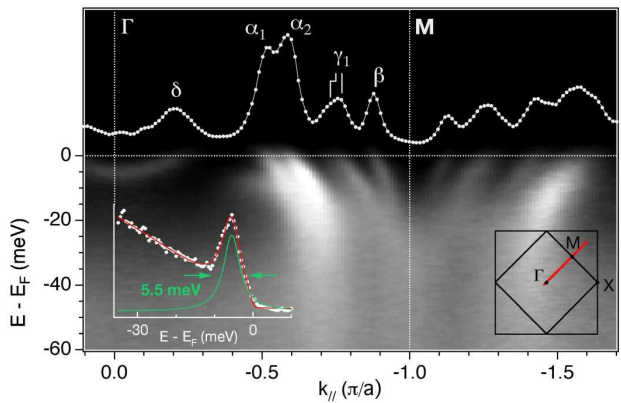}
\caption{\label{f2} Band dispersion along $\Gamma M \Gamma$. The Fermi surface crossings are labeled above the momentum distribution curve (MDC) extracted at $E_F$. The inset shows a spectrum at the $\Gamma$--point. A peak width of 5.5~meV in the raw data indicates an excellent surface quality.
}
\end{figure} 
We start by discussing the Fermi surface of \srII.
A single bilayer of RuO$_6$ octahedra contains 4 Ru$^{4+}$--ions, each contributing 4 conduction electrons distributed over the 3 nearly degenerate $t_{2g}$ levels. Hence, in a first approximation one expects up to 12 bands crossing the Fermi level. Indeed an earlier electronic structure calculation showed a highly fragmented Fermi surface suggesting that its experimental determination could be beyond current capabilities \cite{sin01}. On the other hand, de Haas -- van Alphen experiments found only 5 distinct frequencies corresponding to Fermi surface pockets in the range of 1.1\% to 32\% of the Brillouin zone area \cite{per04, bor04}.

\begin{table*}[htb]
\caption{\label{Rh-table}Fermi surface volumes and cyclotron masses of \srII{} obtained from ARPES and dHvA. The polarity and dominant orbital character of the pockets is indicated in brackets. Errors are estimated from the statistical accuracy of the analysis and the reproducibility of the experiments. The mass of $\gamma_2$ is estimated from parabolic fits to the dispersion.}
\begin{ruledtabular}
\begin{tabular}{l c c c c c c c}
 & $\alpha_1\:(h^+,\;xz,yz)$ &  $\alpha_2\:(h^+,\;xz,yz)$ &  $\beta\:(e^-,\;xz,yz)$ & $\gamma_1\:(e^-,\;xy\;/\;xz,yz )$ & $\gamma_2\:(h^+,\;xy)$ & $\delta\:(e^-,\;x^2-y^2)$ \\
 \hline
ARPES FS--volume $A$ (\% BZ) & 14.1$\pm2$ & 31.5$\pm3$ & 2.6$\pm1$ & 8.0$\pm2$ & $<1$ & 2.1$\pm1$ \\
ARPES cyclotron mass $m^{*}$ ($m_{e}$) & 8.6$\pm3$ & 18$\pm8$ & 4.3$\pm2$ & 9.6$\pm3$ & $10\pm4$ & 8.6$\pm3$\\
\hline
dHvA FS--volume $A$ (\% BZ) & 13.0$\pm$1.0 & 30.1$\pm$1.1 & 1.1$\pm$0.2 & 6.6$\pm0.9$ & -- & 3.1$\pm$0.3 \\
dHvA cyclotron mass $m^{*}$ ($m_{e}$) & 6.9$\pm$0.1 & 10.1$\pm$0.1 & 5.6$\pm$0.3 & 7.7$\pm0.3$ & -- & 8.4$\pm$0.7 \\
\end{tabular}
\end{ruledtabular}
\end{table*}
%
Fig.~1(a) shows the experimental ARPES Fermi surface map. The data have been integrated over $\pm 2$~meV, resulting in an effective energy resolution (convolution of integration window and spectrometer resolution) of 5.5~meV \cite{sym}.
The identification of three Fermi surface pockets centered at $\Gamma$ is straightforward from an analysis of individual cuts. 
The innermost sheet, labeled $\delta$, is nearly circular. Remarkably, its orbital character, as inferred from LDA calculations, is Ru $d_{x^2 - y^2}$, \ie{} it belongs to the $e_g$ manifold, which is unoccupied in \srI{} and \caII{} \cite{mac96ea, dam00ea, bau06aea}.
The larger square- and cross--shaped hole--like Fermi surfaces with pronounced uniaxial anisotropy derive from the out--of--plane $d_{xz,yz}$ orbitals. Around the $M$--points, a small lens ($\beta$) and a larger lens with back--folded vertices ($\gamma_1$) can be identified. The shape of these pockets indicates mixing of $d_{xz,yz}$ and $d_{xy}$ orbital character on the $\gamma_1$ sheet and dominant $d_{xz,yz}$ character for the $\beta$ sheet. The smaller lens is well resolved in the Fermi surface map, while the precise contours of the larger $\gamma_1$ pocket are more difficult to extract. However, its Fermi crossings along the $\Gamma$M$\Gamma$ line can clearly be identified from the cut shown in Fig. 2. A pronounced shoulder on the left hand side of the $\gamma_1$ peak indicates a small bilayer splitting of this pocket, consistent with the DFT calculations. The intensity maxima in Fig.~1(a) connecting the edges of the $\gamma_1$ pockets between adjacent M--points stem from a putative small hole--pocket ($\gamma_2$) that is barely touching the Fermi surface. 
We find the top of this band located at $-1\pm1$~meV (compare Fig. 3). Hence, based on the ARPES data alone we cannot decide unambiguously whether it contributes to the Fermi surface. A slight $k_z$ dispersion or a minute structural difference between surface and bulk might be sufficient to lift the $\gamma_2$ pocket up to the Fermi level for a considerable fraction of the 3D bulk Brillouin zone.

The volumes $A$ of all Fermi surface pockets are summarized in Table 1 and compared with recent de Haas -- van Alphen data \cite{dhva}.
Cyclotron masses $m^*=\frac{\hbar^2}{2\pi}(\partial A/\partial \epsilon)_{k_F}$ have been calculated for individual pockets using Fermi velocities determined along several k--space cuts. 
The agreement is good for the volumes and well within the experimental error for the masses. This strongly suggests that the electronic structure seen by ARPES is nearly converged to that of the bulk. 
Nevertheless, we cannot exclude minute energy differences between surface and bulk, which could lift the $\gamma_2$ pocket up to the Fermi level. To investigate this issue we calculate the specific heat and the Luttinger volume based on the ARPES data \cite{bau06bea}. We tentatively assign multiplicities by assuming a small bilayer splitting of the $\delta$ and $\gamma_1$ pockets, which is unresolved experimentally, and a two-fold degeneracy of $\gamma_2$ as found in the DFT calculations.
With these assumptions we find $\Sigma m^* = 171(36) m_e\: [91(16) m_e]$ corresponding to $\gamma = 127(27)$~mJ/molRuK$^2$ [67(12)~mJ/molRuK$^2$] with [without] counting the $\gamma_2$ pocket. Hence, comparing with the direct measurement of 110~mJ/molRuK$^2$ \cite{ike00} suggests that $\gamma_2$ does contribute to the bulk specific heat and must be included in the zero field Fermi surface.
The above scenario is consistent with the Luttinger volume of a compensated metal with an even number of valence electrons. Assuming an upper bound of 1$\pm 1$\% BZ for $A_{\gamma_2}$ we find $2(-A_{\alpha_1}-A_{\alpha_2}+2A_{\beta}+4A_{\gamma_1}-8A_{\gamma_2}+2A_{\delta}) = -0.24(25)$ electrons/RuO$_2$--bilayer or $-0.06(6)+2n$~electrons/Ru $(n\in \mathbf{N})$.

Note that the high energy resolution and ultra--clean samples employed in this study were instrumental to resolve the Fermi surface of \srII. Some carrier pockets show occupied band widths of only $\approx 5$~meV and Fermi velocities as low as $1.2\cdot10^4$~m/s (0.08~eV\AA), approximately an order of magnitude lower than in single layer \srI. An energy resolution $>10$~meV or impurity scattering contributions of this order to the line width, as are commonly observed by ARPES on transition metal oxides, would render it impossible to resolve such narrow bands. In order to illustrate the present data quality we show an energy distribution curve (EDC) in the inset of Fig. 2 demonstrating a raw quasiparticle line width (including a 4.5~meV contribution from the experimental resolution) of 5.5~meV. This is narrower than state-of-the-art ARPES data from noble metal surface states \cite{rei01} and is rarely seen in complex oxides. 

\begin{figure}[tb]
\includegraphics[width=0.48\textwidth]{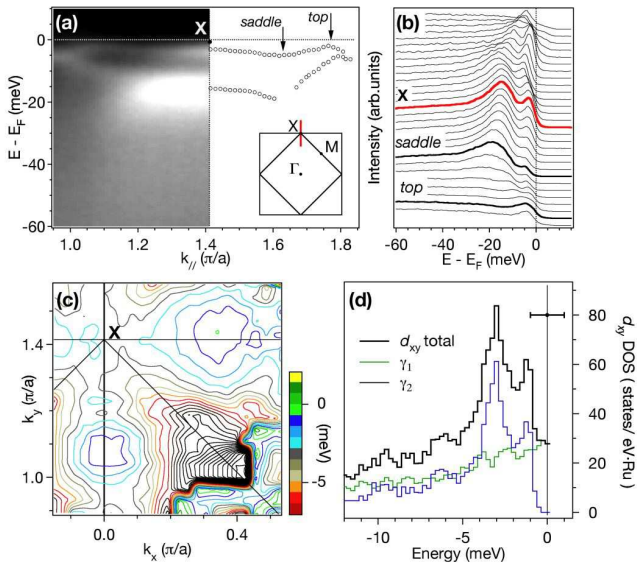}
\caption{\label{f3} Band dispersion around the vHs of the $d_{xy}$ band. (a) Photoemission intensity in the vicinity of  the X point along the direction marked in the inset. Peak positions have been determined by a fit to the energy distribution curves after normalization with the Fermi function. (b) Raw EDCs extracted from the data show in (a). The high intensity above the Fermi level for $k_{\parallel}$ near the top of the band hints at the presence of unoccupied states within a few meV of the Fermi level.
(c) Contour plot showing the energy position of the lowest lying quasiparticle excitation as a function of $(k_{x},k_{y})$. (d) Histogram of near-$E_F$ $k$--states obtained from the band--contours shown in (c). Absolute values are calculated assuming a two-fold degeneracy (or an unresolved bilayer splitting) of $\gamma_{1,2}$. The total density of states $\rho(E_F)$ calculated from the measured specific heat is 47~states/eV/Ru. The horizontal error bar in (d) indicates a $\pm$1~meV uncertainty in absolute $\epsilon(k)$ values estimated from the variation between different measurements.
}
\end{figure} 
The determination of the Fermi surface in agreement with bulk measurements allows for a more in--depth discussion of the electronic structure of \srII{} based on the ARPES data. To this end, we will evaluate the density of states near the Fermi level of the $\gamma_1$, $\gamma_2$ sheets with significant $d_{xy}$ contributions. 
Fig. 3(a) shows the band topography along the $\Gamma$X$\Gamma$ line. A very narrow dispersion with local minima and maxima separated by less than 3~meV is observed over an extended k--space range, indicating a very high DOS just below the Fermi surface. The top of the band corresponding to the putative $\gamma_2$ Fermi surface pocket is observed at $-1\pm1$~meV with a second local maximum at the X--point near $-3$~meV. 
In order to obtain quantitative information on the $d_{xy}$ density of states, we have analyzed the lowest lying excitation at each $k$ point in a high-resolution k--space map covering the relevant area. The resulting contour plot is shown in Fig. 3(c). Saddle points in the dispersion, corresponding to van Hove singularities in the density of states are found symmetrically around the X--point, at an energy of $-4$~meV.
Indeed a histogram of the $\epsilon(k)$ values shows a sharp peak at this energy. A second peak, related to the top of the band is observed around $-1$~meV, coincident with the natural energy scale of metamagnetism.
We emphasize that $\rho(\epsilon)$ obtained in this way is not influenced by matrix-element effects since it is based on counting eigenstates rather than on measuring intensities \cite{dos}. 
Neither is the precision of $\rho(\epsilon)$ limited by the experimental energy resolution of 4.5~meV. It is given by the uncertainty in fitted peak positions, which is dominated by small variations ($<$2~meV) in measured peak positions between different experimental runs.

The sharp peaks in the density of states on the energy scale of the Zeeman splitting shown in Fig.~3(d) is qualitatively similar to the situation considered by Binz and Sigrist in their microscopic model of the metamagnetic transition \cite{bin04}.
This motivates a brief and speculative discussion of the effects of an applied field on the electronic structure of \srII. The pocket most likely to be strongly affected is $\gamma_2$, which could become completely spin polarised in a field of order 10 tesla. Since it is repeated eight times in the BZ each spin polarised pocket would need only to be modest in area to account for the moment change of $\approx0.25\;\mu_B$ at the metamagnetic transition of \srII. The hole-like character of $\gamma_2$ further means that if it dominated the spin polarisation in a rigid-band picture, one would expect an accompanying decrease in the areas of the hole-like sheets $\alpha_1$ and $\alpha_2$ or an increase in the areas of the electron like $\beta$ and $\gamma_1$ sheets. Intriguingly, a pronounced decrease of $\alpha_1$ and $\alpha_2$ was reported in \cite{bor04}, pointing towards an inter--orbital charge transfer from the $d_{xy}$ to the $d_{xz,yz}$ sheet at the metamagnetic transition of \srII. Further investigation of these issues by detailed de Haas -- van Alphen studies and scanning tunnelling spectroscopy in high field are desirable. 

In conclusion, we have presented high--resolution ARPES data from \srII. In combination with new LDA calculations incorporating spin-orbit coupling, the results allow an unambiguous identification of most Fermi surface pockets and of their dominant orbital character. Moreover, our data provide clear evidence for sharp spikes in the quasiparticle density of states on the natural energy scale of metamagnetism.
%
\begin{acknowledgments}
This work has been supported by the Scottish Funding Council and the UK EPSRC. SSRL is operated by the DOE's office of Basic Energy Science. Work at ORNL was supported by DOE BES, Division of Materials Science and Engineering.
\end{acknowledgments}


\begin{thebibliography}{27}
\bibitem{sym}{The orthorhombic crystal structure of \srII{} (space group $Bbcb$ \cite{sha00}) lacks the 4--fold symmetry of \srI{}. However, our LDA calculations show that the Ru--Ru nearest- and next--nearest neighbor directions may be treated as approximate symmetry planes in the band structure. In our experiments, we could not resolve a breaking of these symmetries. Hence, we present the data in Fig.~1(a) symmetrized along $\Gamma$X.}
\bibitem{dos}{Here we assume that matrix elements of all bands in the relevant energy scale are finite and no bands eluded detection.}
\end{thebibliography}

\end{document}